\documentclass[aps,twocolumn,10pt]{revtex4}
\usepackage{graphicx,amssymb,amsmath,subfigure}

\newcommand{\braket}[1]{\left<#1\right>}
\newcommand{\para}[1]{\left(#1\right)}
\newcommand{\abs}[1]{\left|#1\right|}

\begin{document}

\title{Cooperative Electronic and Phononic Mechanism of the High Temperature Superconductivity in Cuprates}
\author{Abolhassan Vaezi}
\affiliation{Department of Physics,
Massachusetts Institute of Technology,
Cambridge, MA 02139, USA}
\email{vaezi@mit.edu}


\begin{abstract}
In conventional superconductors, phonons glue two electrons with opposite spins to form Cooper pairs and condensation of these pairs leads to the superconductivity. Identifying the underlying mechanism of the high temperature superconductivity in cuprates is among the most important problems in physics. Even quarter of a century after the first report of high temperature superconductor by Bednorz and Muller in 1986, there is still no general consensus on the pairing mechanism of superconductivity in these materials. So far, many theories have been developed to explain the exotic properties of cuprates, but they can explain only a limited number of experiments. In this article, we present a new pairing mechanism that incorporates both strong correlation and phonon mediated interaction on an equal footing to produce superconductivity. In this framework, strong correlation and anti-ferromagnetic interaction between electrons, create RVB pairs and phonons provide the phase coherence between these RVB pairs. Both of these are required in this approach to obtain the superconductivity. This approach resolves three limitations of the U(1) slave boson method. We achieve a better estimation of $T_c$, we only predict $\frac{h}{2ec}$ vortices and the linear $T$ coefficient of the superfluid is not sensitive to the doping. This formalism provides a framework that connects Anderson's idea of preformed Cooper pairs  and phonon based theories.
\end{abstract}

\maketitle

\section{Introduction} One scenario for the high temperature superconductivity in cuprates \cite{Bednorz_Mueller_1986} is the preformed Cooper pairs idea that was first proposed by Anderson \cite{Anderson_1987Sci}. This theory is based on the strong correlation effects and it does not incorporate phonons in the pairing mechanism. On the other hand, many experiments have been reported indicating the importance of electron phonon interaction in understanding the physics of high Tc cuprates, such as the oxygen isotope effect on both the transition temperature of superconductivity and the London penetration depth \cite{Vaezi_2010a,Zimmerman_1,Khasanov_3_2008,Raffa_1,Zhao_1,Schneider_Keller_1992_a}.

Observation of the strong isotope effect in cuprates has led many authors to employ the strong limit of the electron phonon interaction as the primary cause of the superconductivity in these materials \cite{Alexandrov_3_1992}. For example, some workers have applied bipolaron theory of superconductivity to the high temperature superconductivity problem. This theory requires the breakdown of the Migdal-Eliashberg theory and is based on the non-adiabatic limit of the electron phonon interaction, where phonons have a much larger energy than electrons. Experimentally, a typical energy of electrons is around the exchange energy $J \sim$ 130 meV \cite{Lee_Nagaosa_Wen_2006a}, while the energy of optical phonons is $40-70$ meV \cite{Iwasawa_1}. On the other hand, the breakdown of the Migdal-Eliashberg theory \cite{Marsiglio_1} in any phonon based theory is crucial, because isotope experiments in cuprates are very different from conventional superconductors that are explained by the BCS theory \cite{BCS} and its generalization, {\em i.e.} Migdal-Eliashberg theory.

Some authors emphasize on the importance of the microscopic inhomogeneity, charge and spin stripes in understanding the mechanism of the superconductivity \cite{Emery_1999a,Kivelson_2003a}. The idea is that the phase segregation can save kinetic energy of holes and the exchange energy of spins. Stripe models are based on the competition between this phase and other states of matter. It is noteworthy that static stripes have been observed only in La$_2$SCuO$_4$ family near $x=\frac{1}{8}$, not in YBCO family, but dynamical stripes are expected even away from $x=\frac{1}{8}$ \cite{PA_Lee_2008_1}.
The idea of Anderson can be implemented using the slave particle method \cite{Lee_Nagaosa_Wen_2006a,Coleman_1983_1,PA_Lee_1992_1,Sachdev_1991_1,Vaezi_2010b}. The U(1) version of this method explains many basic properties of cuprates successfully. However, this method has several limitations. For example, the U(1) slave boson method overestimates the transition temperature of the superconducting state by an order of magnitude. Although in experiments only $\frac{h}{2ec}$ vortices have been observed, this method predicts another $\frac{h}{ec}$ vortices as well. For clean d-wave superconductors, the superfluid density decreases linearly with temperature up to the leading order. Experimentally the coefficient of this linear term is not very sensitive to the doping, while the U(1) slave boson treatment predicts a $x^2$ doping dependence, where $x$ is the doping percentage. In this paper we argue that phonons can mediate attractive interaction between holons and therefore we find a paired holon state. We show that this assumption resolves the three mentioned limitations of the U(1) slave boson approach.

Here we extend the Anderson theory of high $Tc$ which is based on the resonating valence bond(RVB) state, by adding Holstein Hamiltonian such that the model Hamiltonian include the effect of electron phonon interaction. RVB state is the superposition of all possible singlet states between any two sites. The idea of RVB state can be quantified using the slave boson method, and naturally leads to the spin charge separation in two spatial dimensions. The low energy excitations of this state are described by the charged spinless quasiparticles which are called holon and are treated as bosons, and the neutral spin 1/2 quasiparticles which are called spinon and are treated as fermions. This can be implemented by writing the creation operator of physical electrons $c_{i,\sigma}^\dag$ as the product of a fermionic operator $f_{i,\sigma}^\dag$ and a bosonic operator $h_{i}$. Therefore $c_{i,\sigma}^\dag=f_{i,\sigma}^\dag h_{i}$. This assumption comes along with the nondoubly occupancy constraint due to very large onsite repulsion between electrons in cuprates. In terms of slave particles $f_{i,\uparrow}^\dag f_{i,\uparrow}+f_{i,\downarrow}^\dag f_{i,\downarrow}+h_{i}^\dag h_{i}=1$ at every site. Therefore at each site, we have either a holon or a spinon. Using this relation, the density of holon gas is set by doping $x$, and the density of spinons is equal to the density of electrons $1-x$. Spinon pairing forms Cooper pairs in the system and holon condensation provides the phase coherence for these pairs and the superfluid density of the superconducting state is controlled by the condensation fraction of holons. Therefore, to obtain superconducting phase, both holon gas and spinon gas should condense. In case of spinons, we need a pairing mechanism due to their fermionic nature. Strong antiferromagnetic interactions in cuprates can result in such a pairing potential and as a result the transition temperature of spinon paired state is controlled by the exchange energy $J$ which is quite large. This large gap in excitation spectrum of spinons is the origin of the pseudogap phenomenon in cuprates.

On the other hand, due to the bosonic nature of holons, they do not need a pairing mechanism and they can undergo Bose Einstein condensation(BEC). Although this can be in general true, we discuss in the following sections that the phenomenology of cuprates supports the holon pair condensation scenario. This assumption immediately resolves several limitations of the Anderson theory. 

In order to pair holons, like in the BCS theory, we need a pairing mechanism. In our theory, electron phonon interaction can cause such a pairing instability. As we have shown in Ref. \cite{Vaezi_2010a}, this assumption can also explain the observed phonon experiments successfully. In summary in this framework spinons form RVB pairs and holons are also paired by phonons. Both of these are required for the superconductivity in cuprates.

\begin{figure}
  \includegraphics[width=210pt]{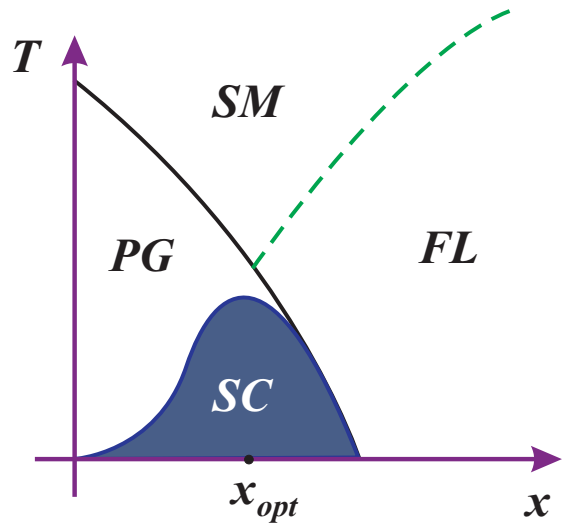}\\
  \caption{ {\bf Phase diagram.---} Schematic phase diagram of cuprates from our theory. $x_{opt}$ is the optimal doping percentage which is typically around \%15. It corresponds to the optimum transition temperature which is typically around 100 Kelvin. SC is the superconducting state in which we obtain spinon pair condensation and holon pair condensation. PG is the pseudogap phase where spinons are paired but holon gas in not condensed. FL is the Fermi liquid phase where we obtain single holon condensation  but no spinon pair condensation. SM is the strange metal in which neither holons nor spinons condense. In this approach, spinons form RVB pairs due to anti-ferromagnetic interactions and holons are paired by phonons. Both of these are required to obtain the superconducting phase in cuprates. We have not shown AF order in this phase diagram, however we have shown in Ref. \cite{Vaezi_2010b} that it can be achieved by adding a nonzero triplet component to the pairing amplitude near half filling.}\label{fig1}
\end{figure}

This approach, gives a phase diagram similar to the phase diagram of the conventional U(1) slave boson treatment of the t-J model, except that the superconducting state has $Z_2$ gauge symmetry, and we obtain a better estimation of $T_c$, closer to the experimental data(see Fig. \ref{fig1}). We have also shown in Ref. \cite{Vaezi_2010b}, that at least at half filling, the antiferromagnetic order in the Hubbard model can be achieved by including a nonzero triplet component in the pairing amplitude besides the singlet component \cite{Vaezi_2010b}. Now let us briefly comment on the experimental consequences of our theory. In appendix, we present more details and derive equations.

\section{Better estimation of $T_c$}
In the underdoped cuprates, both the slave boson theory and the phase fluctuation studies of superconductors predict that the transition temperature of the superconducting state is controlled by the superfluid density. Using Ioffe-Larkin recombination formula (see Appendix C for derivations and discussions), it can be shown that at small dopings, the superfluid density is mostly determined by that of holons. Therefore $T_c$ is determined by the  BEC transition temperature $T_{_{\rm BEC}}$, of holon gas. In the single holon condensation theories we obtain $T^{s}_{_{\rm BEC}}= \frac{2\pi x}{m^{*}_{h}}$. Although this gives a good doping dependence, it is an order of magnitude higher than the superconducting transition temperature \cite{PA_Lee_XG_Wen_1997_1}. In the pair condensation scenario, we obtain the following expression for $T_c$ (see Appendix B for derivation)
\begin{eqnarray}
T^{p}_{_{\rm BEC}}\para{x,\tilde{V}}= \frac{T^{s}_{_{\rm BEC}}}{-\ln \para{2\tilde{V}N(0)}}
\end{eqnarray}
where $\tilde{V}=V\Delta_{s}^2$, where $t\Delta_s$ is the pseudogap energy, is the renormalized coupling constant of the phonon mediated holon holon attraction, and $N(0)=\frac{m^*_{h}}{2\pi }$ is the density of states. In Ref. \cite{Vaezi_2010a}, we had to chose $V=12\frac{\omega_{_E}}{t}$, where $\omega_{_E}$ is the energy of optical phonons, to fit theory and experiment. In cuprates, $12\omega_{_E}\sim t$ and therefore $\tilde{V} \sim \Delta_{s}^{2}\para{T_c}\ll 1$. Because of the extra factor in the denominator, $T^{p}_{_{\rm BEC}}$ is much smaller than $T^{s}_{_{\rm BEC}}$ and is closer to the experimental data.

\section{Vortices}
In the single holon condensation scenario, two kinds of vortices are allowed, $\frac{h}{2ec}$ and $\frac{h}{ec}$. Studies show that the energy of the latter is much smaller, so it should be more stable and visible in experiments. In experiments however, only $\frac{h}{2ec}$ vortices have been observed \cite{Lee_Nagaosa_Wen_2006a}. Absence of $\frac{h}{ec}$ vortices, challenges the assumption of single holon condensation in cuprates. In the holon pair condensation scenario we always obtain $\frac{h}{2ec}$ vortices from both $\left<hh\right>$ or $\left<f_{\uparrow}f_{\downarrow}\right>$ order parameters. This in fact reflects the $Z_2$ structure of our theory.

\section{Linear T coefficient  of the superfluid density}
Another important limitation of the single holon condensation scenario, is the calculation of the linear temperature coefficient of the superfluid density. This scenario predicts a parabolic doping dependence behavior, while experimentally it has a weak dependence on the doping percentage \cite{PA_Lee_XG_Wen_1997_1,PA_Lee_XG_Wen_1998_1}. The reason is that within this assumption, the current carried by quasiparticles is $j=\alpha ev_{_F}$, where $v_{_F}$ is the Fermi velocity of nodal quasiparticles and $\alpha \sim x$ (see Appendix D for details). Since holons are charged particles, they couple to both the external gauge field ($A_{ext}$), and the induced internal gauge field $A_{int}$. Within single holon condensation scenario and using Ioffe-Larkin formula, it can be shown that when both spinons and holons condense, we have $A_{int}\sim -x A_{ext}$. Since spinons are electrically neutral, they only couple to the internal gauge field and therefore they see $-xA_{ext}$, so their effective electric charge is $-xe$. Now we can estimate the value of $\alpha$ by computing the Green's function of real electrons. In the single holon condensation scenario, the diagonal part of the Green's does not respond to the gauge field, because it depends on $\braket{h^\dag}\braket{h}$. Therefore only spinons couple to internal gauge field which is by factor of $-x$ smaller and this is why $\alpha=x$ in this case. In the pair condensation scheme, since $\braket{h}=0$, the diagonal part of holons Green's function depends on $\braket{h_{j}^\dag h_{i}}$. So it responds to the external as well as the internal gauge field. From convolution it can be checked that the real electrons respond to the whole external electromagnetic field and there quasiparticles have effective charge $-e$ (see Appendix D for details). So we obtain $\alpha=1$ in this case, in consistent with the linear temperature coefficient of superfluid density measurements.

In Ref. \cite{Vaezi_2010a}, we have investigated the isotope effect on the superfluid density and the transition temperature of the superconducting phase by considering the effect of electron phonon interaction and have achieved a good agreement between theory and experiment.

It is worth mentioning that boson gas in a purely attractive potential is unstable. It collapses and phase separation happens in that case. However in our model, we deal with holons which are hardcore bosons, {\em i.e.} there is an infinite on-site repulsion between them. The phonon mediated attraction is also screened by the presence of spinons. These two can make the paired state of holons stable. We speculate that under some certain conditions, holons may become meta-stable and form stripes due to the phonon mediated attraction. Moreover as we mentioned earlier, the phase segregation saves the kinetic energy of holes and exchange energy of spinons. This can further stabilize stripe order phase. Therefore, the phonon mediated attraction between holons may enhance the stripe formation.

\section{Conclusion}

In this paper, we have extended the Anderson theory of high $Tc$ to take phonons into consideration. Phonons are engaged in the pairing mechanism by mediating attractive interaction between charged spinless quasiparticles (holons). This attraction destabilizes the single holon condensation state and gives rise to the paired holon state. Assuming the paired holon phase immediately resolves several limitations of the U(1) slave treatment of the Anderson theory. First of all, we achieve a better estimation of the superconducting transition temperature. Moreover, since both spinons and holons are paired, we only find $\frac{h}{2ec}$ vortices. Finally we showed that the effective electric charge of nodal quasiparticles is $-xe$ when we have single holon condensation, so they carry $xev_{_F}$ current, while the the effective electric charge of nodal quasiparticles is $-e$ and they carry $ev_{_F}$ when holons condense in pairs. We have shown that in the latter case, the linear temperature coefficient of the superfluid density is almost independent of doping, consistent with experimental observations.

\section*{ACKNOWLEDGEMENTS}
I thank X.-G. Wen for financial support and very helpful discussions. I acknowledge very useful discussions with T. Senthil, R. Asgari, W. Ketterle, S.-S. Lee B. Swingle, E. Hudson,  M. F. Maghrebi, N. Dehmamy, R. Flint and in particular P.A. Lee. I thank P.W. Anderson for reading preliminary manuscript of this work in its early stages. I acknowledge the clarifying remark by S. Kivelson about the stripe models.

\section*{APPENDICES}

In the following, we provide detailed derivations of formulae in the
main text concerning the expression for $T_c$, Ioffe-Larkin formula and Linear temperature coefficient of superfluid density. In Appendix A, we provide a brief introduction to the slave boson method. In Appendix B, we derive an expression for the transition temperature of the superconducting state. In Appendix C, we derive Ioffe Larkin formula which serves an a powerful tool in relating physical quantities to the corresponding properties of the slave particles. In Appendix D, we present an argument to calculate the linear temperature coefficient of the superfluid density.

\section*{APPENDIX A: METHOD}

Let us start from the t-J model as our starting point. It is believed that this Hamiltonian captures the essential physics of the strongly correlated systems. We finally add the Holstein Hamiltonian to take the effect of electron phonon interaction into consideration. t-J model is defined as the following

\begin{eqnarray}
  H_{t-J}=-t\sum_{\left<i,j\right>, \sigma }P_{G}c_{\sigma,i}^\dag c_{\sigma,j}P_{G}+J\sum_{i,j}\hat{S}_{i}.\hat{S}_{j}
\end{eqnarray}

where $P_{G}$ is the Gutzwiller projection operator that removes doubly occupied states. Within the slave boson formalism, electrons can be decomposed as $c_{i,\sigma}^\dag = f_{i,\sigma}^\dag h_{i} $ along with the physical constraint on each site:  $h_{i}^\dag h_{i}+\sum_{\sigma}f_{i,\sigma}^\dag f_{i,\sigma}=1$ which implements the Gutzwiller projection. $f$ particles are fermions and we call them spinon and $h$ particles are bosons and we call them holon. Spinon corresponds to a state with only one electron and holon to an empty site. The definition of the projected electron operator along with the constraint, says that we have always one slave particle per site and two slave particles cannot sit on the same site. On the other hand, whenever $c_{i,\sigma}^\dag$ acts on an empty site, it annihilates one holon and creates a spinon with spin $\sigma$. We cannot act further on the resulting state by $c_{i,-\sigma}^\dag$, since this operator has to kill a holon, but there is no holon anymore at that site. If we act $c_{i,\sigma}$ on a site that contains a spinon with spin $\sigma$, the operators annihilates the spinon and creates a holon at that site. So by acting projected electron operator we always annihilate one type of slave particle and create another one and therefore the number of slave particles at each site is conserved. Now we can rewrite the t-J model in terms of the new slave particles. Within meanfield approximation and by using Hubbard-Stratonovic transformation, we can decouple spinons (spin sector) from holons (charge sector) and we obtain the following effective Hamiltonians for each sector

\begin{eqnarray}
  &&H_{h}= -\sum_{<i,j>}t\chi_{s} h_{i}^\dag h_{j}-\sum_{i}\mu_{h}h_{i}^\dag h_{i}\\
  &&H_{s}=-\sum_{<i,j>,\sigma}t\chi_{h}f_{i,\sigma}^\dag f_{j,\sigma} -\sum_{i,\sigma}\mu_{s}f_{i,\sigma}^\dag f_{i,\sigma}\cr &&-\sum_{<i,j>}~\para{J/2}\Delta_{s}\para{i,j}\para{f_{i,\uparrow}^\dag f_{j,\downarrow}^\dag-f_{i,\downarrow}^\dag f_{j,\uparrow}^\dag}  +H.c. ,~~
\end{eqnarray}
where the following notations have been used

\begin{eqnarray}
&&  \chi_h=\braket{h_{i+\vec{\delta}}^\dag h_{i}}\\
&&  \chi_s=\braket{\sum_{\sigma}f_{i+\vec{\delta},\sigma}^\dag f_{i,\sigma}}\\
&&  \Delta_{s}\para{i,j}=\frac{1}{2}\braket{f_{i,\uparrow}^\dag f_{j,\downarrow}^\dag-f_{i,\downarrow}^\dag f_{j,\uparrow}^\dag}
\end{eqnarray}

At low temperatures, most of holons occupy the groundstate with momentum $k=0$, therefore $\chi_{h}\sim x$. This model has been extensively studied in the literature and it is well known that this model leads to the d-wave pairing symmetry for spinons \cite{Lee_Nagaosa_Wen_2006a}, {\em i.e.} $\Delta_{s}\para{\pm \hat{x}}=\Delta_{s}$ and $\Delta_{s}\para{\pm \hat{y}}=-\Delta_{s}$.

Now let us consider the electron-phonon interaction. Since the typical energy of electrons is around $J$ and is much larger than $\omega_{_{E}}$, we can apply the BCS theory in our case. Within BCS theory, electron phonon interaction, leads to the following pairing term:

\begin{eqnarray}
  -\sum_{k,k'}V_{k,k'}<c_{k',\uparrow}^\dag c_{-k',\downarrow}^\dag>c_{-k,\downarrow} c_{k,\uparrow}
\end{eqnarray}

By translating the above term to the slave boson language in real space, and using the mean-field approximation, we can substitute $c_{i,\downarrow} c_{j,\uparrow}=f_{i,\downarrow} f_{j,\uparrow}h_{i}^\dag h_{j}^\dag$ by the following terms:

\begin{eqnarray}
\Delta_{h}\para{i,j} f_{i,\downarrow} f_{j,\uparrow} + \Delta_{s}\para{i,j}h_{i}^\dag h_{j}^\dag-\Delta_{s}\para{i,j}\Delta_{h}\para{i,j}
\end{eqnarray}

where $\Delta_{h}\para{i,j}=\braket{h_{i}h_{j}}\sim x$ (doping). Now assuming a very short range interaction, we obtain the following effective interaction:

\begin{eqnarray}
&&  H'_{s-s}=-V \Delta_{h}^2 \sum_{<i,j>}\Delta_{s}\para{i,j}f_{i,\uparrow}^\dag f_{j,\downarrow}^\dag+H.c..\\
&&  H'_{h-h}=-V \Delta_{s}^2 \sum_{<i,j>}\Delta_{h}\para{i,j} h_{i}^\dag h_{j}^\dag +H.c \end{eqnarray}

in which $V=\frac{\gamma_{0}^2}{M\omega_{_E}^2}$ and $\gamma_{0}$ is the bare electron phonon interaction coupling constant, and $\omega_{_E}$ is the energy of optical phonons. Let us assume that $V\Delta_{h}^2 \sim V x^2 \ll J$, so we can neglect this phonon mediated pairing term and therefore, the d-Wave nature of the spinons does not change.  From the above we see that the coupling constant of phonon mediated spinon spinon attraction is renormalized by $\Delta_{h}^2$ factor and that of holons by $\Delta_{s}^2$ due to strong correlation effects. It is easy to show that, $V\propto \gamma^2$, where $\gamma$ is the electron phonon coupling constant. Therefore, we can interpret these renormalization factors as the renormalization of the coupling constant of spinon-phonon interaction to $\Delta_{h}\gamma_{k,q}$ and that of holon-phonon interaction to $\Delta_{s}\gamma_{k,q}$. In Appendix C, using Ioffe-Larkin recombination formula, we present another physical argument to justify this result. Now let us focus on the charge sector (holons). We have studied the effect of holon-phonon interaction on the holon mass in Ref. \cite{Vaezi_2010a}. Since we treat charge sector as a Bose gas, we have shown the holon-phonon interaction is in the non-adiabatic limit and the small polaron picture can be applied. The mass of holons enhances by an exponential factor and we have: $m^{*}_{h}\para{T}=\frac{1}{2\tilde{\chi_{s}}\para{T}}=e^{g^{2}\para{T}}m_{h}$, where  $ g^2\para{T} =\frac{V\Delta_{s}^2\para{T}}{ \omega_{_E}}$.

\section*{APPENDIX B: CALCULATION OF $T_c$}

Taking the mentioned mass renormalization into account, the following continuum model describes the low energy physics of holons:
\begin{eqnarray}
  &&H_{b}=\sum_{k}\para{\epsilon_{k}+\abs{\mu}}h_{k}^\dag h_{k} -\tilde{V} \Delta_{h} \sum_{k} h_{k}^\dag h_{-k} +H.c.~~~~~~
\end{eqnarray}

where $\epsilon_{k}=\frac{k^2}{2m^*_h}$, $\Delta_{h}=\sum_{k}h_{k}h_{-k}$, and $\tilde{V}=V\Delta_{s}^2$. It has been shown that in 2D, no matter how small $\tilde{V}$ is, we have a bound state of bosons and then the double condensation is energetically favorable. Using the Bogoliubov transformation, we can find the energy eigenvalue which is $E_{k}=\sqrt{\para{\epsilon_{k}+\abs{\mu}}^2-\para{\tilde{V}\Delta_{h}}^2}$ and the energy eigenvectors. On the other hand, we should choose $\mu$ such that $x=\frac{1}{N}\sum_{k}h_{k}^\dag h_{k}$. The two mentioned constraints lead to the following self-consistency equations:

\begin{eqnarray}
  &&x=-\frac{1}{2}+\frac{1}{N}\sum_{k}\para{1+2n_{_{BE}}\para{E_{k},T}}\frac{\epsilon_{k}+\abs{\mu}}{2E_{k}}\\
  &&\frac{2}{\tilde{V}}=\frac{1}{N}\sum_{k}\para{1+2n_{_{BE}}\para{E_{k},T}}\frac{1}{2E_{k}}
\end{eqnarray}

in which $n_{_{BE}}\para{E_{k},T}=\frac{1}{\mbox{exp}\para{E_{k}/T}-1}$ is the Bose Einstein distribution function. The first constraint can be solved exactly by $\frac{1}{N}\sum_{k}\rightarrow \frac{1}{2\pi}\int d\vec{k}=N\para{0}\int d \epsilon_{k}$, where $N\para{0}=\frac{2\pi}{m^*_h}$. After integration we obtain:  $E_{g}=\sqrt{\abs{\mu}^2-\para{\tilde{V}\Delta_{h}}^2}=-2T \ln \left\{\frac{\sqrt{4+y^2}-y}{2}\right\}$ where $y=\mbox{exp}\para{\frac{\abs{\mu}-2T_0}{2T}}$ and $T_0=\frac{2\pi x}{m^*_h}$. At $T=0$ we always have $E_g=0$ in consistent with Hugenholtz-Pines theorem. For small values of $\tilde{V}$, we have :

\begin{eqnarray}
  E_g\para{T}=T \mbox{exp}\para{\frac{\abs{\mu}-2T_0}{2T}}
\end{eqnarray}
which is diminishing very rapidly and we can neglect it up to the first order approximation. Therefore, at small enough temperature energy excitations are sound-like an are of the form $E_k=ck$ where $c=\sqrt{\frac{\abs{\mu}}{m^*_h}}$. At $T_{c}^{p}$, $\Delta_b=0$ and therefore $E_k=\epsilon_{k}+\abs{\mu}$ and $E_{g}=\abs{\mu}=T_c^p \mbox{exp}\para{\frac{\abs{\mu}-2T_0}{2T_c^p}}\simeq T_c^p \mbox{exp}\para{-\frac{T_0}{T_c^p}}$. The second constraint can be written in the following way:

\begin{eqnarray}
  &&\frac{4}{N\para{0}\tilde{V}}=\int \frac{\coth\para{\frac{\epsilon+\abs{\mu}}{2T_c^p}}}{\epsilon+\abs{\mu}}d\epsilon \\
  &&\frac{4}{N\para{0}\tilde{V}}\simeq \int \frac{2T_c^p}{\para{\epsilon+\abs{\mu}}^2}d\epsilon \simeq \frac{2T_c^p}{\abs{\mu}}=2\mbox{exp}\para{\frac{T_0}{T_c^p}}~~~~~\\
  && T_c^p=\frac{T_0}{-\ln 2N\para{0}\tilde{V}}=\frac{2\pi x}{-m^*_h \ln \left\{\frac{4\pi V\Delta_{s}^2}{m^{*}_{h}}\right\}}
\end{eqnarray}

Using the paired state and assuming gapless excitations ($E_g =0$), we can show that up to the first order, $\braket{h_{i,t}^\dag h_{j,0}}=x_{pc}$, where $x_{pc}$ is the ground-state macroscopic occupation number (condensation fraction), though $\braket{h_{i,t}}=0$. If we replace every operator by minus itself the effective Hamiltonian and all order parameters remain the same, and therefore we have $Z_2$ gauge freedom and the low energy theory is described by a $Z_2$ gauge theory. As long as $T<T_c^p$, the condensation fraction is nonzero and $T=0$, $x_{pc}=x$, where $x$ is doping (number of holons).

\section*{APPENDIX C: IOFFE-LARKIN FORMULA} One important question that should be addressed is how to relate the physical quantities to the corresponding quantities of spinons and holons? For example, given the conductivity of spinons and holons, what is the conductivity of real electrons? To answer this question we should note two things. First of all, since at each site we should have one slave particle, therefore if one spinons hops from site $i$ to site $j$, it should be accompanied by a hopping of one holons from site $j$ to site $i$. So we conclude that the current carried by spinons is equal but opposite to the current carried by holons and they add up to zero. Since electron operator at site $i$ with spin $\sigma$, is written in terms of slave particles as $c_{i,\sigma}^\dag=f_{i,\sigma}^\dag h_{i}$, it is invariant under U(1) gauge transformation, provided spinons and holons carry the same charge under this transformation. On the other hand, we assume spinons to be electrically neutral and assign $+e$ electric charge to holons (we could also assume neutral holons and assign $-e$ electric charge to spinons). By scaling the internal gauge field we can assume $e_{int}=e$. As we discussed we should satisfy the following constraint

\begin{eqnarray}
  \vec{J}_{s}+\vec{J}_{h}=0  \end{eqnarray}\label{Eq20}

Since only holons carry electric charge we have

\begin{eqnarray}
  \vec{J}_{ph}=\vec{J}_{h}=-\vec{J}_{s}  \end{eqnarray}

Above two equations tells us that nonzero external gauge field (electromagnetic field) induces internal gauge field. Because we have

\begin{eqnarray}
  &&\vec{J}_{s}=e\sigma_{s}E_{int}\\
  &&\vec{J}_{h}=e\sigma_{h}\para{E_{ext}+E_{int}}
\end{eqnarray}

Solving Eq. [19]

\begin{eqnarray}
  &&E_{int}=-\frac{\sigma_{h}}{\sigma_{s}+\sigma_{h}}E_{ext}
\end{eqnarray}

and as a result $\vec{J}_{ph}=e\frac{\sigma_{h}\sigma_{s}}{\sigma_{s}+\sigma_{h}}E_{ext}$. Equivalently

\begin{eqnarray}
  \sigma_{ph}=\frac{\sigma_{h}\sigma_{s}}{\sigma_{s}+\sigma_{h}}
\end{eqnarray}

Now let us consider two important cases. Fermi liquid phase and the Superconducting phase.

\subsection{\bf Fermi Liquid Phase} In the Fermi liquid (FL) phase, holons are condensed but spinons are not. So we have $\braket{h}\neq 0$ and $\Delta_{s}=0$. Since in this case, $\sigma_{h} \gg \sigma_{s}$, we have $E_{int}\simeq-E_{ext}$, $\sigma_{ph}\simeq \sigma_{s}$, and $\vec{J}_{ph}\simeq e\sigma_{s}E_{ext}$. Now let us define the effective electric charge of spinons and holons as

\begin{eqnarray}
  \vec{J}_{ph}= -e_{s}\sigma_{s}E_{ext}=e_{h}\sigma_{h}E_{ext}
\end{eqnarray}

Therefore in the Fermi liquid we have $e_{s}\simeq -e$ and therefore they have nonzero overlap with physical electrons and carry the same charge. On the other hand $e_{h}\simeq 0$ and we can safely assume that holons are electrically neutral in the FL phase. This can bee directly seen from $c_{i,\sigma}^\dag \simeq \braket{h} f_{i,\sigma}^\dag$. One important result of this argument is that in the absence of the pseudogap ({\em i.e.} when $\Delta_s=0$), condensed holons do not couple to phonons, since phonons only couple to electrically charged quasiparticles. In other words, phonons create local electromagnetic field and this field induces another (internal) gauge field. Holons couple to the sum of these two fields. When $\Delta_s=0$, but holons condense ({\em i.e.} at low enough temperatures) these two fields cancel out each other and as a result holons do not couple to phonons.

\subsection{\bf Superconducting Phase} In the superconducting (SC) phase , both holons and spinons are condensed. So we have $\Delta_{h}=\braket{hh}\neq 0$ and $\Delta_{s}\neq 0$. In this case, $\frac{\sigma_{h}}{\sigma_{s}}=\frac{\rho_{c,h}}{\rho_{c,s}}$, where $\rho_{c,h}$ and $\rho_{c,s}$ is the condensation fraction of holon and spinon gas respectively. At zero temperature all holons and spinons condense and therefore $\frac{\sigma_{h}}{\sigma_{s}}=\frac{x}{1-x}$ and we have $E_{int}\simeq-xE_{ext}$, $\sigma_{ph}\simeq x\sigma_{s}$, and $\vec{J}_{ph}\simeq xe\sigma_{s}E_{ext}$. We can compute the effective charge of spinons and holons and we obtain $e_{s}\simeq -xe$ and $e_{h}\simeq \para{1-x}e$ respectively. Therefore in the superconducting state, spinons only respond to the $x$ fraction of the electromagnetic field. Since holons are effectively charged quasiparticles in this state, (or since the internal gauge field does not cancel out the local electromagnetic field of phonons completely) they couple to phonons. The larger pseudogap value, the stronger interaction between holons and phonons. This is an intuitive way to justify our recipe for renormalization of the electron phonon coupling constant from the brae value $\gamma$, to the effective value $\Delta_s \gamma$. Using this expression and since $\Delta_{s}$ decreases with doping, we can easily explain why the isotope effect is a decreasing function of doping as well. Moreover, in the overdoped region, $\Delta_{s}=0$ at the transition temperature and therefore holons do not interact with phonons and as a result we do not expect isotope effect on the $T_c$. However at $T=0$, $\Delta_{s}\neq 0$ even in the overdoped region and the mass of holons enhances in an isotope dependent way and this explains the nonvanishing isotope effect on the superfluid density and the London penetration depth in this region.

\section*{APPENDIX D: LINEAR T COEFFICIENT OF THE SUPERFLUID DENSITY} Another important limitation of the single holon condensation scenario, is the calculation of the linear temperature coefficient of the superfluid density. This scenario predicts a parabolic doping dependence behavior, while experimentally it has a weak dependence on the doping percentage \cite{PA_Lee_XG_Wen_1997_1,PA_Lee_XG_Wen_1998_1}. The reason is that within this assumption, the current carried by quasiparticles is $j=\alpha ev_{_F}$, where $v_{_F}$ is the Fermi velocity of nodal quasiparticles and $\alpha \sim x$. Lee and Wen have shown that the linear temperature dependence of a d-Wave superconductor is given by the following expression

\begin{eqnarray}
  \frac{\rho_{s}\para{T}}{m}=\frac{x}{m}-\frac{2\ln2}{\pi}\alpha^2 \para{\frac{v_{_F}}{v_2}}T
\end{eqnarray}

where $v_2$ is the velocity of the d-wave SC quasiparticles in the direction perpendicular to $v_{_F}$. Now let us give a simple argument on how to compute $\alpha$. Since holons are charged particles, they couple to both the external gauge field ($A_{ext}$), and the induced internal gauge field $A_{int}$. Within single holon condensation scenario and using Ioffe-Larkin formula, it can be shown that when both spinons and holons condense, we have $A_{int}\sim -x A_{ext}$. Since spinons are electrically neutral, they only couple to the internal gauge field and therefore they see $-xA_{ext}$, so their effective electric charge is $-xe$. Now we can estimate the value of $\alpha$ by computing the Green's function of real electrons. Green's function can be calculated by convoluting the Green's function of holons with that of spinons. Within the single holon condensation scenario $g_{e}\para{k+eA_{ex}/c,\omega}=x_{c} g_{s}\para{k+eA_{int}/c,\omega}$, where $x_c$ is the condensation fraction of holon gas. This expression means that the physical quasiparticles only response to the induced internal field $A_{int}\sim -x A_{ext}$, and therefore their effective electric charge is $-xe$. Therefore they carry $xev_{_F}$ current and this leads to $\alpha=x$. On the other hand, in the double condensation scenario, as it has been discussed by Lee and Wen \cite{PA_Lee_XG_Wen_1997_1,PA_Lee_XG_Wen_1998_1}, we have $g_{e}\para{k+eA_{ext}/c,\omega}=x_{pc}g_{s}\para{k+eA_{ext}/c,\omega}$. Therefore, in this case, quasiparticles see the whole external field and we obtain $\alpha=1$ in consistent with the linear temperature coefficient of superfluid density measurements. Let us do more serious calculations now. In the pair condensed scenario, since $x_{pc}\para{T}$ holons per lattice site condense at the ground state with energy $E_g$ (which we showed before is exponentially small), we can write the diagonal part of the holons Green's function as

\begin{eqnarray}
  ig_{h}\para{k,\omega}=x_{pc}\delta\para{k}\delta\para{\omega-E_{g}}+i\tilde{g}_{h}\para{k,\omega}
\end{eqnarray}
where $\tilde{g}_{h}$ denotes the uncondensed part of the system. The Green's function of spinons is
\begin{eqnarray}
  &&ig_{s}\para{k,\omega}=\frac{\abs{u_{k}}^2}{\omega+E_{s,k}-i0^+}-\frac{\abs{v_{k}}^2}{\omega-E_{s,k}-i0^+}~~~\\
  &&i\Delta_{s}\para{k,\omega}=\frac{u_{k}v_{k}}{\omega+E_{s,k}-i0^+}-\frac{u_{k}v_{k}}{\omega-E_{s,k}-i0^+}~~~
\end{eqnarray}

Now let us choose the Coulomb gauge in which $A_{0}=0$ and $\nabla.\vec{A}=0$. Since holons are charged quasiparticles, they couple to the both internal and external gauge fields. Since we have assumed pair condensation, the diagonal part of Green's function responds to the gauge fields and we have
\begin{eqnarray}
  &&ig_{A,h}\para{k,\omega}=g_{h}\para{k-e\para{A_{int}+A_{ext}}/c,\omega}
\end{eqnarray}

In the presence of the gauge field only the diagonal part of the spinons Green's function responds to the gauge field and the off-diagonal part does not change. Since spinons are electrically neutral, they only couple to the internal gauge field
\begin{eqnarray}
  &&g_{A,s}\para{k,\omega}=g_{s}\para{k-eA_{int}/c,\omega}
\end{eqnarray}

From $c_{i,\sigma}^\dag=f_{i,\sigma}^\dag h_{i}$ it can be read that the Green's function of the real electrons is related to the Green's function of holons and spinons by convolution.

\begin{eqnarray}
  g_{A,e}\para{k,\omega}=i\int\frac{d^2Q d\Omega }{\para{2\pi}^3} g_{A,s}\para{k+Q,\omega+\Omega}g_{A,h}\para{Q,\Omega}~~~
\end{eqnarray}

Let us separate the coherent and incoherent parts of the Green's function $g_{e}\para{k,\omega}=g^{coh}_{e}\para{k,\omega}+g^{inc}_{e}\para{k,\omega}$. For the coherent part we immediately conclude

\begin{eqnarray}
  &&g^{coh}_{A,e}\para{k,\omega}=g^{coh}_{e}\para{k+eA_{ext}/c,\omega}
\end{eqnarray}

Which clearly implies that the effective charge of quasiparticles is $-e$ and therefore they carry $ev_{_F}$ current, not $xev_{_F}$. So we conclude that in the pair condensation scenario, quasiparticles carry the whole current and therefore $\alpha=1$ in this case.

In the single condensation scenario, the diagonal part of the holons Green's function is also $ig_{h}\para{k,\omega}=\abs{\braket{h}}^2\delta\para{k}\delta\para{\omega}+i\tilde{g}_{h}\para{k,\omega}$. However the first term does not respond to the gauge field, like the offdiagonal part of the electrons Green's function and we have  $ig_{A,h}\para{k,\omega}=\abs{\braket{h}}^2\delta\para{k}\delta\para{\omega}+i\tilde{g}_{h}\para{k-e\para{A_{int}+A_{ext}}/c,\omega}$. After convolution it can be verified that the coherent part of the real electrons Green's function is $g^{coh}_{A,e}\para{k,\omega}=g^{coh}_{e}\para{k-eA_{int}/c,\omega}$. In the previous section we showed that in the SC state and at very low temperatures, $A_{int}\simeq-xA_{ext}$, so we have $g^{coh}_{A,e}\para{k,\omega}=g^{coh}_{e}\para{k+xeA_{ext}/c,\omega}$. This results $\alpha=x$ and therefore quasiparticles carry only $x$ fraction of the $ev_{_F}$ current. So the single holon condensation scenario gives $x^2$ dependence for the linear temperature dependence of the superfluid density, which is far from experimental observations.


\begin{thebibliography}{22}
\expandafter\ifx\csname natexlab\endcsname\relax\def\natexlab#1{#1}\fi
\expandafter\ifx\csname bibnamefont\endcsname\relax
  \def\bibnamefont#1{#1}\fi
\expandafter\ifx\csname bibfnamefont\endcsname\relax
  \def\bibfnamefont#1{#1}\fi
\expandafter\ifx\csname citenamefont\endcsname\relax
  \def\citenamefont#1{#1}\fi
\expandafter\ifx\csname url\endcsname\relax
  \def\url#1{\texttt{#1}}\fi
\expandafter\ifx\csname urlprefix\endcsname\relax\def\urlprefix{URL }\fi
\providecommand{\bibinfo}[2]{#2}
\providecommand{\eprint}[2][]{\url{#2}}

\bibitem[{\citenamefont{{Bednorz} and {Mueller}}(1986)}]{Bednorz_Mueller_1986}
\bibinfo{author}{\bibfnamefont{J.~G.} \bibnamefont{{Bednorz}}}
  \bibnamefont{and} \bibinfo{author}{\bibfnamefont{K.~A.}
  \bibnamefont{{Mueller}}}, \bibinfo{journal}{Z. Phys. B64}
  \textbf{\bibinfo{volume}{64}}, \bibinfo{pages}{189} (\bibinfo{year}{1986}).

\bibitem[{\citenamefont{{Anderson}}(1987)}]{Anderson_1987Sci}
\bibinfo{author}{\bibfnamefont{P.~W.} \bibnamefont{{Anderson}}},
  \bibinfo{journal}{Science} \textbf{\bibinfo{volume}{235}},
  \bibinfo{pages}{1196} (\bibinfo{year}{1987}).

\bibitem[{\citenamefont{{Vaezi}}(2010)}]{Vaezi_2010a}
\bibinfo{author}{\bibfnamefont{A.}~\bibnamefont{{Vaezi}}},
  \bibinfo{journal}{arXiv e-prints}  (\bibinfo{year}{2010}),
  \eprint{arXiv:1009.4721}.

\bibitem[{\citenamefont{Zimmermann et~al.}(1995)\citenamefont{Zimmermann,
  Keller, Lee, Savi\ifmmode~\acute{c}\else \'{c}\fi{}, Warden, Zech, Cubitt,
  Forgan, Kaldis, Karpinski et~al.}}]{Zimmerman_1}
\bibinfo{author}{\bibfnamefont{P.}~\bibnamefont{Zimmermann}},
  \bibinfo{author}{\bibfnamefont{H.}~\bibnamefont{Keller}},
  \bibinfo{author}{\bibfnamefont{S.~L.} \bibnamefont{Lee}},
  \bibinfo{author}{\bibfnamefont{I.~M.}
  \bibnamefont{Savi\ifmmode~\acute{c}\else \'{c}\fi{}}},
  \bibinfo{author}{\bibfnamefont{M.}~\bibnamefont{Warden}},
  \bibinfo{author}{\bibfnamefont{D.}~\bibnamefont{Zech}},
  \bibinfo{author}{\bibfnamefont{R.}~\bibnamefont{Cubitt}},
  \bibinfo{author}{\bibfnamefont{E.~M.} \bibnamefont{Forgan}},
  \bibinfo{author}{\bibfnamefont{E.}~\bibnamefont{Kaldis}},
  \bibinfo{author}{\bibfnamefont{J.}~\bibnamefont{Karpinski}},
  \bibnamefont{et~al.}, \bibinfo{journal}{Phys. Rev. B}
  \textbf{\bibinfo{volume}{52}}, \bibinfo{pages}{541} (\bibinfo{year}{1995}).

\bibitem[{\citenamefont{Khasanov et~al.}(2008)\citenamefont{Khasanov,
  Shengelaya, Di~Castro, Morenzoni, Maisuradze, Savi\ifmmode~\acute{c}\else
  \'{c}\fi{}, Conder, Pomjakushina, Bussmann-Holder, and
  Keller}}]{Khasanov_3_2008}
\bibinfo{author}{\bibfnamefont{R.}~\bibnamefont{Khasanov}},
  \bibinfo{author}{\bibfnamefont{A.}~\bibnamefont{Shengelaya}},
  \bibinfo{author}{\bibfnamefont{D.}~\bibnamefont{Di~Castro}},
  \bibinfo{author}{\bibfnamefont{E.}~\bibnamefont{Morenzoni}},
  \bibinfo{author}{\bibfnamefont{A.}~\bibnamefont{Maisuradze}},
  \bibinfo{author}{\bibfnamefont{I.~M.}
  \bibnamefont{Savi\ifmmode~\acute{c}\else \'{c}\fi{}}},
  \bibinfo{author}{\bibfnamefont{K.}~\bibnamefont{Conder}},
  \bibinfo{author}{\bibfnamefont{E.}~\bibnamefont{Pomjakushina}},
  \bibinfo{author}{\bibfnamefont{A.}~\bibnamefont{Bussmann-Holder}},
  \bibnamefont{and} \bibinfo{author}{\bibfnamefont{H.}~\bibnamefont{Keller}},
  \bibinfo{journal}{Phys. Rev. Lett.} \textbf{\bibinfo{volume}{101}},
  \bibinfo{pages}{077001} (\bibinfo{year}{2008}).

\bibitem[{\citenamefont{Raffa et~al.}(1998)\citenamefont{Raffa, Ohno, Mali,
  Roos, Brinkmann, Conder, and Eremin}}]{Raffa_1}
\bibinfo{author}{\bibfnamefont{F.}~\bibnamefont{Raffa}},
  \bibinfo{author}{\bibfnamefont{T.}~\bibnamefont{Ohno}},
  \bibinfo{author}{\bibfnamefont{M.}~\bibnamefont{Mali}},
  \bibinfo{author}{\bibfnamefont{J.}~\bibnamefont{Roos}},
  \bibinfo{author}{\bibfnamefont{D.}~\bibnamefont{Brinkmann}},
  \bibinfo{author}{\bibfnamefont{K.}~\bibnamefont{Conder}}, \bibnamefont{and}
  \bibinfo{author}{\bibfnamefont{M.}~\bibnamefont{Eremin}},
  \bibinfo{journal}{Phys. Rev. Lett.} \textbf{\bibinfo{volume}{81}},
  \bibinfo{pages}{5912} (\bibinfo{year}{1998}).

\bibitem[{\citenamefont{{Zhao} et~al.}(1997)\citenamefont{{Zhao}, {Hunt},
  {Keller}, and {M{\"u}ller}}}]{Zhao_1}
\bibinfo{author}{\bibfnamefont{G.}~\bibnamefont{{Zhao}}},
  \bibinfo{author}{\bibfnamefont{M.~B.} \bibnamefont{{Hunt}}},
  \bibinfo{author}{\bibfnamefont{H.}~\bibnamefont{{Keller}}}, \bibnamefont{and}
  \bibinfo{author}{\bibfnamefont{K.~A.} \bibnamefont{{M{\"u}ller}}},
  \bibinfo{journal}{Nature} \textbf{\bibinfo{volume}{385}},
  \bibinfo{pages}{236} (\bibinfo{year}{1997}).

\bibitem[{\citenamefont{Schneider and Keller}(1992)}]{Schneider_Keller_1992_a}
\bibinfo{author}{\bibfnamefont{T.}~\bibnamefont{Schneider}} \bibnamefont{and}
  \bibinfo{author}{\bibfnamefont{H.}~\bibnamefont{Keller}},
  \bibinfo{journal}{Phys. Rev. Lett.} \textbf{\bibinfo{volume}{69}},
  \bibinfo{pages}{3374} (\bibinfo{year}{1992}).

\bibitem[{\citenamefont{Alexandrov}(1992)}]{Alexandrov_3_1992}
\bibinfo{author}{\bibfnamefont{A.~S.} \bibnamefont{Alexandrov}},
  \bibinfo{journal}{Phys. Rev. B} \textbf{\bibinfo{volume}{46}},
  \bibinfo{pages}{14932} (\bibinfo{year}{1992}).

\bibitem[{\citenamefont{{Lee} et~al.}(2006)\citenamefont{{Lee}, {Nagaosa}, and
  {Wen}}}]{Lee_Nagaosa_Wen_2006a}
\bibinfo{author}{\bibfnamefont{P.~A.} \bibnamefont{{Lee}}},
  \bibinfo{author}{\bibfnamefont{N.}~\bibnamefont{{Nagaosa}}},
  \bibnamefont{and} \bibinfo{author}{\bibfnamefont{X.}~\bibnamefont{{Wen}}},
  \bibinfo{journal}{Reviews of Modern Physics} \textbf{\bibinfo{volume}{78}},
  \bibinfo{pages}{17} (\bibinfo{year}{2006}).

\bibitem[{\citenamefont{Iwasawa et~al.}(2008)\citenamefont{Iwasawa, Douglas,
  Sato, Masui, Yoshida, Sun, Eisaki, Bando, Ino, Arita et~al.}}]{Iwasawa_1}
\bibinfo{author}{\bibfnamefont{H.}~\bibnamefont{Iwasawa}},
  \bibinfo{author}{\bibfnamefont{J.~F.} \bibnamefont{Douglas}},
  \bibinfo{author}{\bibfnamefont{K.}~\bibnamefont{Sato}},
  \bibinfo{author}{\bibfnamefont{T.}~\bibnamefont{Masui}},
  \bibinfo{author}{\bibfnamefont{Y.}~\bibnamefont{Yoshida}},
  \bibinfo{author}{\bibfnamefont{Z.}~\bibnamefont{Sun}},
  \bibinfo{author}{\bibfnamefont{H.}~\bibnamefont{Eisaki}},
  \bibinfo{author}{\bibfnamefont{H.}~\bibnamefont{Bando}},
  \bibinfo{author}{\bibfnamefont{A.}~\bibnamefont{Ino}},
  \bibinfo{author}{\bibfnamefont{M.}~\bibnamefont{Arita}},
  \bibnamefont{et~al.}, \bibinfo{journal}{Phys. Rev. Lett.}
  \textbf{\bibinfo{volume}{101}}, \bibinfo{pages}{157005}
  (\bibinfo{year}{2008}).

\bibitem[{\citenamefont{{Marsiglio} and {Carbotte}}(2001)}]{Marsiglio_1}
\bibinfo{author}{\bibfnamefont{F.}~\bibnamefont{{Marsiglio}}} \bibnamefont{and}
  \bibinfo{author}{\bibfnamefont{J.~P.} \bibnamefont{{Carbotte}}},
  \bibinfo{journal}{arXiv Condensed Matter e-prints}  (\bibinfo{year}{2001}),
  \eprint{arXiv:cond-mat/0106143}.

\bibitem[{\citenamefont{Bardeen et~al.}(1957)\citenamefont{Bardeen, Cooper, and
  Schrieffer}}]{BCS}
\bibinfo{author}{\bibfnamefont{J.}~\bibnamefont{Bardeen}},
  \bibinfo{author}{\bibfnamefont{L.~N.} \bibnamefont{Cooper}},
  \bibnamefont{and} \bibinfo{author}{\bibfnamefont{J.~R.}
  \bibnamefont{Schrieffer}}, \bibinfo{journal}{Phys. Rev.}
  \textbf{\bibinfo{volume}{108}}, \bibinfo{pages}{1175} (\bibinfo{year}{1957}).

\bibitem[{\citenamefont{{Emery} et~al.}(1999)\citenamefont{{Emery}, {Kivelson},
  and {Tranquada}}}]{Emery_1999a}
\bibinfo{author}{\bibfnamefont{V.~J.} \bibnamefont{{Emery}}},
  \bibinfo{author}{\bibfnamefont{S.~A.} \bibnamefont{{Kivelson}}},
  \bibnamefont{and} \bibinfo{author}{\bibfnamefont{J.~M.}
  \bibnamefont{{Tranquada}}}, \bibinfo{journal}{Proceedings of the National
  Academy of Science} \textbf{\bibinfo{volume}{96}}, \bibinfo{pages}{8814}
  (\bibinfo{year}{1999}), \eprint{arXiv:cond-mat/9907228}.

\bibitem[{\citenamefont{Kivelson et~al.}(2003)\citenamefont{Kivelson, Bindloss,
  Fradkin, Oganesyan, Tranquada, Kapitulnik, and Howald}}]{Kivelson_2003a}
\bibinfo{author}{\bibfnamefont{S.~A.} \bibnamefont{Kivelson}},
  \bibinfo{author}{\bibfnamefont{I.~P.} \bibnamefont{Bindloss}},
  \bibinfo{author}{\bibfnamefont{E.}~\bibnamefont{Fradkin}},
  \bibinfo{author}{\bibfnamefont{V.}~\bibnamefont{Oganesyan}},
  \bibinfo{author}{\bibfnamefont{J.~M.} \bibnamefont{Tranquada}},
  \bibinfo{author}{\bibfnamefont{A.}~\bibnamefont{Kapitulnik}},
  \bibnamefont{and} \bibinfo{author}{\bibfnamefont{C.}~\bibnamefont{Howald}},
  \bibinfo{journal}{Rev. Mod. Phys.} \textbf{\bibinfo{volume}{75}},
  \bibinfo{pages}{1201} (\bibinfo{year}{2003}).

\bibitem[{\citenamefont{{Lee}}(2008)}]{PA_Lee_2008_1}
\bibinfo{author}{\bibfnamefont{P.~A.} \bibnamefont{{Lee}}},
  \bibinfo{journal}{Reports on Progress in Physics}
  \textbf{\bibinfo{volume}{71}}, \bibinfo{pages}{012501}
  (\bibinfo{year}{2008}), \eprint{0708.2115}.

\bibitem[{\citenamefont{Coleman}(1983)}]{Coleman_1983_1}
\bibinfo{author}{\bibfnamefont{P.}~\bibnamefont{Coleman}},
  \bibinfo{journal}{Phys. Rev. B} \textbf{\bibinfo{volume}{28}},
  \bibinfo{pages}{5255} (\bibinfo{year}{1983}).

\bibitem[{\citenamefont{Lee and Nagaosa}(1992)}]{PA_Lee_1992_1}
\bibinfo{author}{\bibfnamefont{P.~A.} \bibnamefont{Lee}} \bibnamefont{and}
  \bibinfo{author}{\bibfnamefont{N.}~\bibnamefont{Nagaosa}},
  \bibinfo{journal}{Phys. Rev. B} \textbf{\bibinfo{volume}{46}},
  \bibinfo{pages}{5621} (\bibinfo{year}{1992}).

\bibitem[{\citenamefont{Read and Sachdev}(1991)}]{Sachdev_1991_1}
\bibinfo{author}{\bibfnamefont{N.}~\bibnamefont{Read}} \bibnamefont{and}
  \bibinfo{author}{\bibfnamefont{S.}~\bibnamefont{Sachdev}},
  \bibinfo{journal}{Phys. Rev. Lett.} \textbf{\bibinfo{volume}{66}},
  \bibinfo{pages}{1773} (\bibinfo{year}{1991}).

\bibitem[{\citenamefont{{Vaezi} and {Wen}}(2010)}]{Vaezi_2010b}
\bibinfo{author}{\bibfnamefont{A.}~\bibnamefont{{Vaezi}}} \bibnamefont{and}
  \bibinfo{author}{\bibfnamefont{X.}~\bibnamefont{{Wen}}},
  \bibinfo{journal}{arXiv e-prints}  (\bibinfo{year}{2010}),
  \eprint{arXiv:1010.5744}.

\bibitem[{\citenamefont{Lee and Wen}(1997)}]{PA_Lee_XG_Wen_1997_1}
\bibinfo{author}{\bibfnamefont{P.~A.} \bibnamefont{Lee}} \bibnamefont{and}
  \bibinfo{author}{\bibfnamefont{X.-G.} \bibnamefont{Wen}},
  \bibinfo{journal}{Phys. Rev. Lett.} \textbf{\bibinfo{volume}{78}},
  \bibinfo{pages}{4111} (\bibinfo{year}{1997}).

\bibitem[{\citenamefont{Wen and Lee}(1998)}]{PA_Lee_XG_Wen_1998_1}
\bibinfo{author}{\bibfnamefont{X.-G.} \bibnamefont{Wen}} \bibnamefont{and}
  \bibinfo{author}{\bibfnamefont{P.~A.} \bibnamefont{Lee}},
  \bibinfo{journal}{Phys. Rev. Lett.} \textbf{\bibinfo{volume}{80}},
  \bibinfo{pages}{2193} (\bibinfo{year}{1998}).

\end{thebibliography}
\end{document}